\documentclass[times,twocolumn]{aastex61} %
\usepackage{graphicx}
\usepackage{natbib}   %

\usepackage{url}
\def\spose#1{\hbox to 0pt{#1\hss}}
\newcommand\lsim{\mathrel{\spose{\lower 3pt\hbox{$\mathchar"218$}}
     \raise 2.0pt\hbox{$\mathchar"13C$}}}
\newcommand\gsim{\mathrel{\spose{\lower 3pt\hbox{$\mathchar"218$}}
     \raise 2.0pt\hbox{$\mathchar"13E$}}}
\def\ltsima{$\; \buildrel < \over \sim \;$}
\def\lsim{\lower.5ex\hbox{\ltsima}}
\def\gtsima{$\; \buildrel > \over \sim \;$}
\def\gsim{\lower.5ex\hbox{\gtsima}}

\def\apjl{ApJL}
\def\apjs{ApJSS}

\def\aap{A\&A}

\shorttitle{The June 2015 $\gamma$-ray flare of 3C 279}
\shortauthors{Pittori C., et al.}

\begin{document}

\title{The bright $\gamma$-ray flare of 3C 279 in June 2015: AGILE detection and multifrequency follow-up observations}

\correspondingauthor{Carlotta Pittori}
\email{carlotta.pittori@ssdc.asi.it}

\author{C. Pittori}
\affiliation{\centering
ASI Space Science Data Center (SSDC), Via del Politecnico snc, I-00133 Roma, Italy}
\affiliation{\centering
INAF, Osservatorio Astronomico di Roma, via Frascati 33, I-00078 Monte Porzio Catone (Roma), Italy}

\author{F. Lucarelli}
\affiliation{\centering
ASI Space Science Data Center (SSDC), Via del Politecnico snc, I-00133 Roma, Italy}
\affiliation{\centering
INAF, Osservatorio Astronomico di Roma, via Frascati 33, I-00078 Monte Porzio Catone (Roma), Italy}

\author{F. Verrecchia}
\affiliation{\centering
ASI Space Science Data Center (SSDC), Via del Politecnico snc, I-00133 Roma, Italy}
\affiliation{\centering
INAF, Osservatorio Astronomico di Roma, via Frascati 33, I-00078 Monte Porzio Catone (Roma), Italy}

\author{C. M. Raiteri}
\affiliation{\centering
INAF, Osservatorio Astrofisico di Torino, via Osservatorio 20, I-10025 Pino Torinese, Italy}

\author{M. Villata}
\affiliation{\centering
INAF, Osservatorio Astrofisico di Torino, via Osservatorio 20, I-10025 Pino Torinese, Italy}

\author{V. Vittorini}
\affiliation{\centering
INAF/IAPS--Roma, Via del Fosso del Cavaliere 100, I-00133 Roma, Italy}

\author{M. Tavani}
\affiliation{\centering
INAF/IAPS--Roma, Via del Fosso del Cavaliere 100, I-00133 Roma, Italy}
\affiliation{\centering
Univ. ``Tor Vergata'', Via della Ricerca Scientifica 1, I-00133 Roma, Italy}
\affiliation{\centering
Gran Sasso Science Institute, viale Francesco Crispi 7, I-67100 L'Aquila, Italy}

\author{S. Puccetti}
\affiliation{\centering
Agenzia Spaziale Italiana (ASI), Via del Politecnico snc, I-00133 Roma, Italy}

\author{M. Perri}
\affiliation{\centering
ASI Space Science Data Center (SSDC), Via del Politecnico snc, I-00133 Roma, Italy}
\affiliation{\centering
INAF--OAR, via Frascati 33, I-00078 Monte Porzio Catone (Roma), Italy}

\author{I. Donnarumma}
\affiliation{\centering
Agenzia Spaziale Italiana (ASI), Via del Politecnico snc, I-00133 Roma, Italy}
\affiliation{\centering
INAF/IAPS--Roma, Via del Fosso del Cavaliere 100, I-00133 Roma, Italy}

\author{S. Vercellone}
\affiliation{\centering
INAF, Osservatorio Astronomico di Brera, Via E. Bianchi 46, I-23807 Merate (LC), Italy}

\author{J. A. Acosta-Pulido}
\affiliation{\centering Instituto de Astrofisica de Canarias (IAC), E-38205 La Laguna, Tenerife, Spain}
\affiliation{\centering Departamento de Astrofisica, Universidad de La Laguna, E-38206 La Laguna, Tenerife, Spain}

\author{R. Bachev}
\affiliation{\centering Institute of Astronomy and National Astronomical Observatory, Bulgarian Academy of Sciences, 72 Tsarigradsko shosse Blvd., 1784 Sofia, Bulgaria}

\author{E. Ben\'itez}
\affiliation{\centering Instituto de Astronom\'a, Universidad Nacional Aut\'onoma de M\'exico, Apdo. Postal 70-264, 04510 Cd. de M\'exico, Mexico}

\author{G. A. Borman}
\affiliation{\centering Crimean Astrophysical Observatory, P/O Nauchny, 298409, Russia}

\author{M. I. Carnerero}      
\affiliation{\centering INAF, Osservatorio Astrofisico di Torino, via Osservatorio 20, I-10025 Pino Torinese, Italy}

\author{D. Carosati}      
\affiliation{\centering EPT Observatories, Tijarafe, E-38780 La Palma, Spain}
\affiliation{\centering INAF, TNG Fundaci\'on Galileo Galilei, E-38712 La Palma, Spain}

\author{W. P. Chen}
\affiliation{\centering Graduate Institute of Astronomy, National Central University, Jhongli City, Taoyuan County 32001, Taiwan}

\author{Sh. A. Ehgamberdiev}
\affiliation{\centering Ulugh Beg Astronomical Institute, Maidanak Observatory, 33 Astronomicheskaya str., Tashkent, 100052, Uzbekistan} 

\author{A. Goded}
\affiliation{\centering Instituto de Astrofisica de Canarias (IAC), E-38205 La Laguna, Tenerife, Spain}
\affiliation{\centering Departamento de Astrofisica, Universidad de La Laguna, E-38206 La Laguna, Tenerife, Spain}

\author{T. S. Grishina}
\affiliation{\centering Astronomical Institute, St.\ Petersburg State University, 198504 St.\ Petersburg, Russia}

\author{D. Hiriart}
\affiliation{\centering Instituto de Astronomía, Universidad Nacional Aut\'onoma de M\'exico, Ensenada, Baja California, Mexico}

\author{H. Y. Hsiao}
\affiliation{\centering Graduate Institute of Astronomy, National Central University, Jhongli City, Taoyuan County 32001, Taiwan}

\author{S. G. Jorstad}
\affiliation{\centering Institute for Astrophysical Research, Boston University, 725 Commonwealth Avenue, Boston, MA 02215, USA}
\affiliation{\centering Astronomical Institute, St. Petersburg State University, Universitetskij Pr. 28, Petrodvorets,
198504 St. Petersburg, Russia}

\author{G. N. Kimeridze}
\affiliation{\centering Abastumani Observatory, Mt. Kanobili, 0301 Abastumani, Georgia}

\author{E. N. Kopatskaya}
\affiliation{\centering Astronomical Institute, St.\ Petersburg State University, 198504 St.\ Petersburg, Russia}

\author{O. M. Kurtanidze}
\affiliation{\centering Abastumani Observatory, Mt. Kanobili, 0301 Abastumani, Georgia}
\affiliation{\centering Engelhardt Astronomical Observatory, Kazan Federal University, Tatarstan,
Russia}

\author{S. O. Kurtanidze}
\affiliation{\centering Abastumani Observatory, Mt. Kanobili, 0301 Abastumani, Georgia}

\author{V. M. Larionov}
\affiliation{\centering Astronomical Institute, St.\ Petersburg State University, 198504 St.\ Petersburg, Russia}
\affiliation{\centering Pulkovo Observatory, 196140 St.\ Petersburg, Russia}

\author{L. V. Larionova}
\affiliation{\centering Astronomical Institute, St.\ Petersburg State University, 198504 St.\ Petersburg, Russia}

\author{A. P. Marscher}
\affiliation{\centering Institute for Astrophysical Research, Boston University, 725 Commonwealth Avenue, Boston, MA 02215, USA}

\author{D. O. Mirzaqulov}
\affiliation{\centering Ulugh Beg Astronomical Institute, Maidanak Observatory, 33 Astronomicheskaya str., Tashkent, 100052, Uzbekistan}

\author{D. A. Morozova}
\affiliation{\centering Astronomical Institute, St.\ Petersburg State University, 198504 St.\ Petersburg, Russia}

\author{K. Nilsson}
\affiliation{\centering Finnish Centre for Astronomy with ESO (FINCA), University of Turku, V\"ais\"al\"antie 20, FI-21500 Piikki\"o, Finland}

\author{M. R. Samal}
\affiliation{\centering Graduate Institute of Astronomy, National Central University, Jhongli City, Taoyuan County 32001, Taiwan}

\author{L. A. Sigua}
\affiliation{\centering Abastumani Observatory, Mt. Kanobili, 0301 Abastumani, Georgia}

\author{B. Spassov}
\affiliation{\centering Institute of Astronomy and National Astronomical Observatory, Bulgarian Academy of Sciences, 72 Tsarigradsko shosse Blvd., 1784 Sofia, Bulgaria}

\author{A. Strigachev}
\affiliation{\centering Institute of Astronomy and National Astronomical Observatory, Bulgarian Academy of Sciences, 72 Tsarigradsko shosse Blvd., 1784 Sofia, Bulgaria}

\author{L.O. Takalo}
\affiliation{\centering Tuorla Observatory, Department of Physics and Astronomy, University of Turku, FI-20014 Turku, Finland}

\author{L. A. Antonelli}
\affiliation{\centering
ASI Space Science Data Center (SSDC), Via del Politecnico snc, I-00133 Roma, Italy}

\author{A. Bulgarelli}
\affiliation{\centering
INAF-IASF Bologna, via Gobetti 101, I-40129 Bologna, Italy}

\author{P. Cattaneo}
\affiliation{\centering
INFN-Pavia, via Bassi 6, I-27100 Pavia, Italy
}
\author{S. Colafrancesco}
\affiliation{\centering
School of Physics, University of the Witwatersrand, Johannesburg Wits 2050, South Africa}

\author{P. Giommi}
\affiliation{\centering
Agenzia Spaziale Italiana (ASI), Via del Politecnico snc, I-00133 Roma, Italy}

\author{F. Longo}
\affiliation{\centering
Dipartimento di Fisica, Univ. di Trieste and INFN, via Valerio 2, I-34127 Trieste, Italy
}

\author{A. Morselli}
\affiliation{\centering
INFN Roma ``Tor Vergata'', via della Ricerca Scientica 1, I-00133 Roma, Italy
}

\author{F. Paoletti}
\affiliation{\centering
East Windsor RSD, 25a Leshin Lane, Hightstown, NJ 08520, USA}
\affiliation{\centering
INAF/IAPS--Roma, Via del Fosso del Cavaliere 100, I-00133 Roma, Italy}

\vskip 0.5 truecm

\begin{abstract}
We report the AGILE detection and the results of the multifrequency follow-up observations of a
bright $\gamma$-ray flare of the blazar 3C 279 in June 2015. We use AGILE-GRID and
{\it Fermi}-LAT $\gamma$-ray data, together with
{\it Swift}-XRT, {\it Swift}-UVOT, and ground-based GASP-WEBT optical
observations, including polarization information, to study the
source variability and the overall spectral energy distribution during the $\gamma$-ray flare.
The $\gamma$-ray flaring data, compared with as yet unpublished
simultaneous optical data which allow to set constraints on the
big blue bump disk luminosity, show  very high Compton
dominance values of $\sim 100$, with a ratio of $\gamma$-ray to
optical emission rising by a factor of three in a few hours. The
multi-wavelength behavior of the source during the flare
challenges one-zone leptonic theoretical models. The
new observations during the June 2015 flare are also compared with
already published data and non-simultaneous historical 3C 279
archival data.
\end{abstract}



\section{Introduction}
\label{sec-intro}
Blazars are a subclass of radio-loud active galactic nuclei (AGN) with relativistic jets
pointing towards the observer~\citep{1995PASP..107..803U}. 
Their emission extends from the radio band to the 
$\gamma$-ray band above 100 MeV, up to TeV $\gamma$-rays, 
and it is dominated by variable non-thermal processes. They
come in two main flavors, with very different optical spectra:
Flat Spectrum Radio Quasars (FSRQs) which have strong, broad
optical emission lines, and BL Lacertae objects (BL Lacs) with an
optical spectrum which can be completely featureless, or can show
at most weak emission lines and some absorption features
(e.g., see \citealt{2012MNRAS.420.2899G} for a detailed review on blazar classification).
The blazar spectral energy distribution (SED) is in general characterized by two
broad bumps: a low-energy one, spanning from the radio to the
X-ray band, is attributed to synchrotron radiation, while the
high-energy one, from the X-ray to the $\gamma$-ray band, is
thought to be due to inverse Compton (IC) emission.
In the leptonic scenario this second component is due to
relativistic energetic electrons scattering their own synchrotron
photons (Synchrotron self-Compton, SSC) or  photons external to
the jet (External Compton, EC).
Blazars of both flavors have been found to be highly variable, and particularly so in
$\gamma$-rays\footnote{SED movie of the 
blazar 3C279 from 2008.05 to 2016.37 by P. Giommi: https://www.youtube.com/watch?v=o0lJBakFUXQ}. 
Correlated variability between X-rays and $\gamma$-rays is usually
well explained in the SSC or EC framework~\citep{1998MNRAS.301..451G}. 
In fact, a new class of \lq \lq orphan'' $\gamma$-ray flares from FSRQ blazars is
now emerging from observations, challenging the current simple
one-zone leptonic models. 
In particular, a number of $\gamma$-ray flares from some
extensively monitored FSRQs such as 3C 279 do not correlate with optical and soft X-ray events of comparable power and time scales, see for example the
results of a previous multi-wavelength campaign on 3C 279 during flaring states in
2013-2014~\citep{2015ApJ...807...79H}. 

Gamma-ray observations of flaring blazars and simultaneous multi-wavelength data
are thus the key to investigate possible alternative theoretical scenarios,
such as a recently proposed model based on a
mirror-driven process within a
clumpy jet inducing localized and transient enhancements of synchrotron photon
density beyond the broad-line region (BLR)~\citep{2015ApJ...814...51T,Vittorini2017}.
Other scenarios consider special structures,
such as spine-sheath jet layers radiative interplay
~\citep{Tavecchio2008,Sikora2016}, 
or \lq \lq rings'' of fire, i.e. synchrotron-emitting rings of
electrons representing a shocked portion of the jet sheath~\citep{MacDonald2015}.

3C 279 is associated with a luminous FSRQ at z = 0.536~\citep{1965ApJ...142.1667L} with prominent broad emission lines detected in all accessible spectral bands, and revealing highly variable emission. It consistently shows strong $\gamma$-ray
emission, already clearly detected by EGRET \citep{1992ApJ...385L...1H,Kniffen1993},
AGILE \citep{2009A&A...494..509G}, Fermi-LAT \citep{Hayashida2012,2015ApJ...807...79H},
and also detected above 100 GeV by MAGIC \citep{Albert2008}.
The central black hole
mass estimates are in the range of $(3 - 8)\times 10^8 M_\odot$~\citep{2001MNRAS.327.1111G,2002ApJ...579..530W,2009A&A...505..601N}.
The 3C279 jet features strings of compact plasmoids 
as indicated by radio observations \citep{Hovatta2009}, 
which may be a by-product of the magnetic reconnection process \citep{Petropoulou2016}, even though it must be taken into account that the superluminal knots observed in Very Long Baseline Interferometry (VLBI) images are probably much larger structures 
than reconnection plasmoids formed on kinetic plasma scales, 
hence this connection is uncertain \citep{2008ApJ...689...79C}. 

Here we present the results of a multi-band observing campaign on the blazar
3C 279 triggered by the detection of intense $\gamma$-ray emission above
100 MeV by the AGILE satellite in June 2015~\citep{2015ATel.7631....1L}.
The source is one of the $\gamma$-ray blazars monitored by the GLAST-AGILE Support Program (GASP) of the Whole Earth Blazar Telescope (WEBT) Collaboration\footnote{http://www.oato.inaf.it/blazars/webt/}~\citep{Villata2008,Bottcher2007,Larionov2008,Abdo2010}.

AGILE-GRID $\gamma$-ray data of 3C 279 in June 2015 are compared with
as yet unpublished ({\it R}-band) optical GASP-WEBT observations during the
flare, including percentage and angle of polarization, and with
{\it Fermi}-LAT~\citep{Ackermann2016,2015ApJ...803...15P} and
other multi-wavelength data from {\it {\it Swift}}-UVOT and {\it {\it Swift}}-XRT
Target of Opportunities (ToOs).
The analysis of the source multi-wavelength behavior is crucial in
order to  study the correlation, if any, of the $\gamma$-ray
radiation with the optical-UV and X-ray emissions.
The June 2015 flaring data are also compared
with non-simultaneous archival data from the NASA/IPAC Extragalactic Database (NED)
and from the ASI Space Science Data Center (SSDC, previously known as ASDC).

\section{Observations and Data Analysis}
\label{sec-data}

\begin{table*}[t!]
\centering
\begin{tabular}{ccccc}  
\hline
\hline
~~~OBS Start Time &  ~~~~~~~MJD & XRT exposure & UVOT exposure & ~~~obsID \\  
 ~~~~~~~(UTC)   & ~~~~~~~ & ~~~~~~~ (s)  & ~~~~~~ (s)  & \\ %
\hline
2015-06-15 14:27:58  & ~~~~57188.6028 & ~~~~~ 1987.8  & ~~~ 1994.1  & 00035019176 \\  
2015-06-16 03:27:59  & ~~~~57189.1444 & ~~~~~~ 958.9  & ~~~~ 961.7  & 00035019180 \\   
2015-06-16 16:04:58  & ~~~~57189.6701 & ~~~~~~ 934.0  & ~~~~ 936.1  & 00035019181 \\   
2015-06-17 04:40:59  & ~~~~57190.1951 & ~~~~~~ 936.5  & ~~~~ 938.2  & 00035019185 \\   
2015-06-17 20:59:58  & ~~~~57190.8750 & ~~~~~~ 489.5  & ~~~~ 488.6  & 00035019187 \\   
2015-06-18 04:37:59  & ~~~~57191.1930 & ~~~~~ 1246.1  & ~~~ 1249.2  & 00035019188 \\  
\hline
\hline
\end{tabular}
\caption{{\it Swift} follow-up observations of 3C 279 following the
AGILE $\gamma$-ray flare alert in June 2015, and
on-source net exposures in pointing observing mode
for the XRT (Photon Counting readout mode) and UVOT instruments
within each observation.
}
\label{swift-summary}
\end{table*}

\subsection{AGILE observations} 
AGILE~\citep{2009A&A...502..995T} is a small mission of the Italian Space Agency (ASI) devoted to $\gamma$-ray
astrophysics, operating in a low Earth orbit since April 23, 2007.
The main AGILE instrument is the Gamma-Ray Imaging Detector
(GRID), which is sensitive in the energy range 30 MeV -- 50 GeV.
The AGILE-GRID consists of a silicon-tungsten tracker, a caesium iodide mini-calorimeter (MCAL), and an anticoincidence system (AC) made of segmented plastic scintillators.

The AGILE Quick Look (QL) alert system~\citep{2013NuPhS.239..104P,2014ApJ...781...19B}
detected increased $\gamma$-ray emission from 3C 279 starting from 2015, June 13 (MJD=57186)
which lasted up to 2015, June 17 (MJD=57190).

AGILE-GRID data were analyzed using the AGILE Standard Analysis Pipeline (see \citealt{Vercellone2008} 
for a description of the AGILE data reduction).
Counts, exposure and Galactic
diffuse background maps for energy E $\geq 100$ MeV were created including all events collected up to 60$^{\circ}$ off--axis.
Scientific data acquisition is inhibited during the
South Atlantic Anomaly (SAA) passages, and we rejected all $\gamma$-ray events whose
reconstructed directions form angles with the satellite-Earth vector $\le 80^{\circ}$
to reduce the $\gamma$-ray Earth albedo contamination.
We used the latest public AGILE software Package (AGILE SW 5.0 SourceCode) with Calibration files (I0023), and the AGILE $\gamma$-ray diffuse emission model~\citep{2004MSAIS...5..135G} publicly available at the SSDC site\footnote{http://agile.ssdc.asi.it/publicsoftware.html}.

\subsection{GASP-WEBT Observations}
Optical observations of 3C 279 were carried out by the GASP-WEBT Collaboration in the Cousins' $R$ band. Data were provided by the following observatories: Abastumani (Georgia), Belogradchik (Bulgaria),  Crimean (Russia), Lowell (USA; Perkins telescope), Lulin (Taiwan), Mt. Maidanak (Uzbekistan), Roque de los Muchachos (Spain; KVA), San Pedro Martir (Mexico), Skinakas (Greece), St. Petersburg (Russia), Teide (Spain; IAC80 and STELLA-I), and Tijarafe (Spain).
The calibrated source magnitude was obtained by differential photometry with respect to Stars 1, 2, 3, and 5 of the photometric sequence by \citet{Raiteri1998}. The optical light curve (see Sect. \ref{sect-lc}) was visually inspected and checked. No significant offset was noticed between different datasets.
Polarimetric information in the $R$ band 
was acquired at the Crimean, Lowell, San Pedro Martir, and St. Petersburg observatories.

\subsection{{\it Swift} ToO observations}
Following the 3C 279 $\gamma$-ray flare detected by AGILE,
a prompt {\it {\it Swift}} target of opportunity observation 
was performed on 2015, June 15, for a total net exposure time of about 2.0 ks.
Other five 
{\it Swift}-XRT observations were carried out on 2015, June 16--18.
A summary of these observations is given in Table \ref{swift-summary}, where the net exposures with the XRT and UVOT instruments are also reported.

\subsubsection{XRT observations}
The XRT on board {\it Swift} \citep{2004ApJ...611.1005G} is sensitive to the 0.3--10~keV X-ray energy band~\citep{2004SPIE.5165..201B}. The six 
2015 June XRT follow-up observations of 3C 279 were all carried out using the most sensitive Photon Counting (PC) readout mode for a total net exposure time of about
6.5 ks. 
The XRT data sets were first processed with the XRTDAS software package (v.3.1.0) developed at SSDC and distributed by
HEASARC within the HEASoft package (v. 6.17). Event files were calibrated and cleaned with standard filtering criteria
with the {\it xrtpipeline} task using the calibration files available in the version 20150721 of the {\it Swift}-XRT CALDB.
Except for 
the last two observations, the source count rate was initially high enough to cause some photon pile-up in the inner 3 pixels radius circle centered on the source position,
as derived from the comparison of the observed PSF profile with the analytical
model derived in \cite{2005SPIE.5898..360M}. 
We avoided pile-up effects by selecting events within an annular region with an inner radius of 3 pixels and an outer radius of 30 pixels. The background was extracted from a nearby source-free annular region of 50/90 pixel inner/outer radius.
The ancillary response files were generated with the {\it xrtmkarf} task, applying corrections for the PSF losses and CCD defects using the cumulative exposure map.  The response matrices available in the {\it Swift} CALDB at the time of analysis were used.
The source spectra were binned to ensure a minimum of 30 counts per bin.

For all {\it Swift} ToO observations, fits of the XRT spectra were performed
using the XSPEC package.
The observed X-ray spectrum (0.3--10~keV) can be fit by an absorbed power-law model with a HI column density consistent
with the Galactic value in the direction of the source, $n_H = 2.2 \times 10^{20}$ cm$^{-2}$~\citep{2005A&A...440..775K}.
The results of photon index and 
fluxes corrected for the Galactic absorption 
for each follow-up observation are shown in Table \ref{xspec}.

\begin{table}[t]
\centering
\centerline{3C 279}
\vskip 0.1 truecm
\begin{tabular}{lll}
\hline
\hline
~~XRT Date Start   &  photon index & ~~~ XRT Flux (0.3-10 keV)    \\  
 ~~~~~~~~(UTC)  & ~~~~          & ~~~~~ (erg cm$^{-2}$ s$^{-1}$)   \\
\hline
2015-06-15 14:32   & ~~1.36$\pm$0.06  & ~~~~$(5.5 \pm 0.4) \times 10^{-11}$  \\
2015-06-16 03:31   & ~~1.32$\pm$0.08  & ~~~~$(9.4 \pm 0.8) \times 10^{-11}$  \\
2015-06-16 16:08   & ~~1.4$\pm$0.1    & ~~~~$(3.5 \pm 0.5) \times 10^{-11}$  \\
2015-06-17 04:44  & ~~1.4$\pm$0.1    & ~~~~$(2.7 \pm 0.4) \times 10^{-11}$  \\
2015-06-17 21:02   & ~~1.3$\pm$0.2    & ~~~~$(2.0 \pm 0.5) \times 10^{-11}$  \\
2015-06-18 04:41   & ~~1.5$\pm$0.1    & ~~~~$(1.7 \pm 0.2) \times 10^{-11}$  \\
\hline
\hline
\end{tabular}
\caption{Results of the X-ray spectral analysis of the {\it Swift}-XRT follow up data.
The errors are at 90\% level of confidence, and fluxes are corrected for the Galactic absorption.
}
\label{xspec}
\end{table}

\vskip 0.5 cm
\subsubsection{UVOT observations}
Co-aligned with the X-Ray Telescope, the {\it Swift}-UVOT instrument \citep{2005SSRv..120...95R}
provides simultaneous ultraviolet and optical coverage ($170-650$ nm).
UVOT ToO observations were performed with the optical/UV filter of the day,
namely U, W2 and M2, as described in Table \ref{UVOT}.
We performed aperture photometry using the standard UVOT software distributed within the HEAsoft package (version 6.17)
and the calibration included in the latest release of the CALDB.
The values of the UVOT observed magnitudes of the source are given in Table~\ref{UVOT}.
Source counts were extracted from aperture of 5 arcsec radius for all filters, while the background ones from
an annular region of inner aperture 26 arcsec and size 9 arcsec, then the source counts were converted to
fluxes using the standard zero points \citep{Breeveld2011}.
The fluxes were finally de-reddened using the appropriate value of $E(B-V)=0.0245$ taken
from \citet{1998ApJ...500..525S} and \citet{2011ApJ...737..103S},
with $A_{\lambda}/E(B-V)$ ratios calculated for UVOT filters using the mean Galactic interstellar
extinction curve from \citet{1999PASP..111...63F}.
These fluxes were then included in the multi-wavelength SED
(see Sect. \ref{sect-sed}).

\begin{table}[t]
\centering
\centerline{3C 279}
\vskip 0.1 truecm
\begin{tabular}{lll}  
\hline
\hline
~~ UVOT Date Start   &  ~~~ Filter    & UVOT Magnitude    \\  
  ~~~~~~ (UTC)       &     (of the day)  & ~~~~~   \\ 
\hline
2015-06-15 14:33  & ~~~~~~ U ~~~~ & $ 14.93 \pm 0.03 $  \\
2015-06-16 03:32  & ~~~~~ W2 ~~~~ & $ 15.35 \pm 0.04 $  \\
2015-06-16 16:09  & ~~~~~ W2 ~~~~ & $ 15.44 \pm 0.04 $  \\
2015-06-17 04:45  & ~~~~~ M2 ~~~~ & $ 15.38 \pm 0.04 $  \\
2015-06-17 21:04  & ~~~~~ M2 ~~~~ & $ 15.64 \pm 0.05 $  \\
2015-06-18 04:41  & ~~~~~ W1 ~~~~ & $ 15.65 \pm 0.04 $  \\
\hline
\hline
\end{tabular}
\caption{Results of the analysis of the {\it Swift}-UVOT ToO follow up data. 
Observed magnitudes, not corrected for Galactic extinction,
and errors at $1 \sigma$ confidence level.
}
\label{UVOT}
\end{table}

\subsection{{\it Fermi}-LAT observations}
We compared AGILE $\gamma$-ray observations with
published {\it Fermi}-LAT data from \citet{Ackermann2016}, 
and with public Fermi data obtained from the online data analysis 
tool at SSDC\footnote{https://tools.asdc.asi.it/?\&searchtype=fermi}.
As described in \citet{Ackermann2016},
events in the energy range 100 MeV--300 GeV were extracted within a $15^{\circ}$ acceptance cone of the Region of Interest (ROI) centered on the location of the source. Gamma-ray fluxes and spectra were determined by an unbinned maximum likelihood fit with \texttt{gtlike}. The background model included all known $\gamma$-ray sources 
within the ROI from the 3rd {\it Fermi}-LAT catalog \citep{Acero2015}.
Additionally, the model included the isotropic and Galactic diffuse emission components. Flux normalization for the diffuse and background sources were left free in the fitting procedure.

\begin{figure*}[h!]
\centering
\includegraphics[width=15cm]{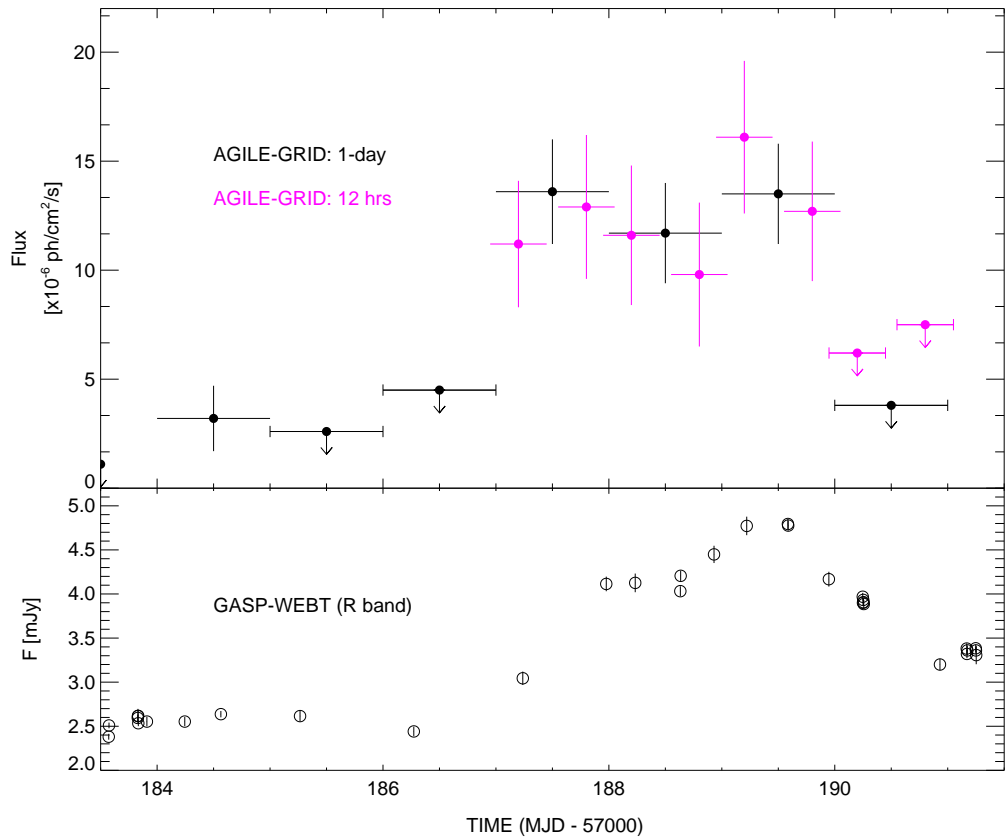}
\vskip -3 truecm
\caption{Upper panel: AGILE-GRID 3C279 $\gamma$-ray 
light curve ($E \geq 100$ MeV) during the June 2015 flare.
Lower panel: simultaneous GASP-WEBT optical data
({\it R}-band, de-absorbed flux densities), showing
a well-defined maximum peaking around MJD=57189.
}
\label{GRID-GASP}
\end{figure*}
\begin{figure*}
\centering
\includegraphics[width=15cm]{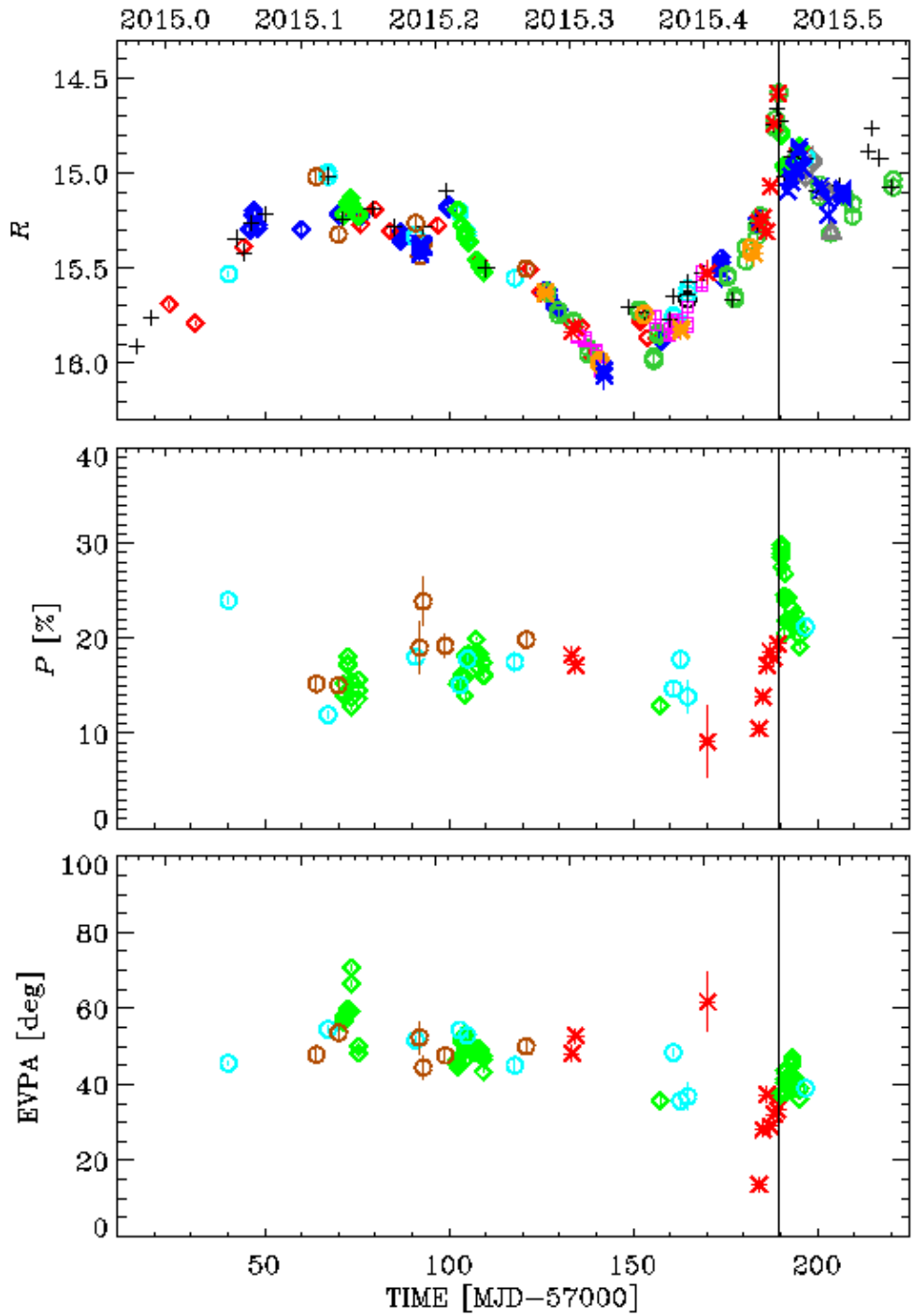}
\caption{
Photometric and polarimetric optical data in the $R$ band acquired by the GASP-WEBT Collaboration from 2014, December 9 (MJD=57000) to 2015, July 17 (MJD=57220). 
The corresponding time in calendar years is shown above the figure.
Different colors and symbols highlight data points from different telescopes (see text for the full list). The vertical line indicates the optical flux measured maximum (MJD=57189.585).
}
\label{GASP-WEBT}
\end{figure*}

\begin{figure*}[h!]
\vskip 0.2 true cm
\epsscale{1.0}
\plotone{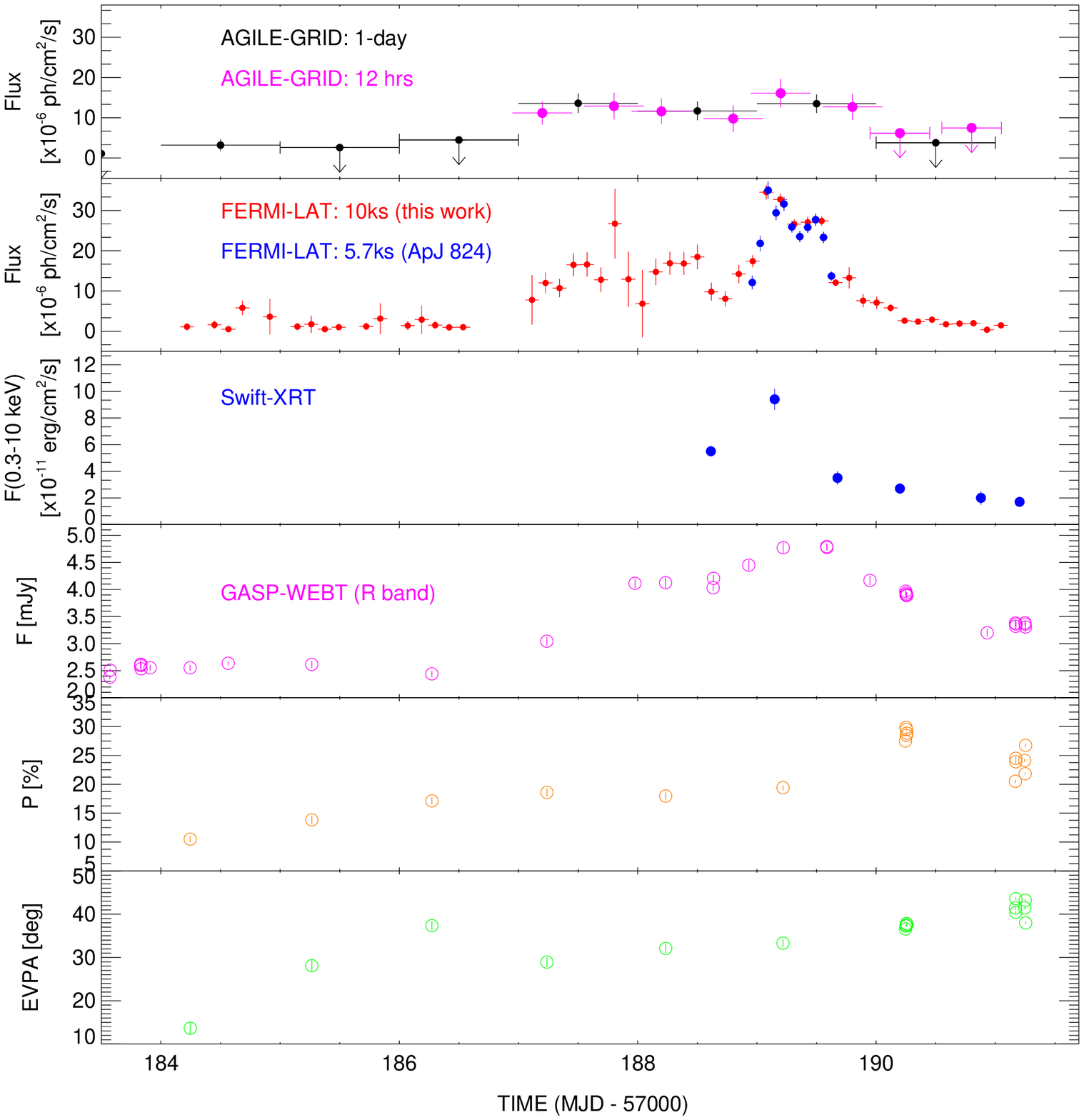}
\caption{Multi-wavelength light curves of 3C 279 in June 2015: $\gamma$-rays ($E \geq 100$ MeV) as observed by AGILE-GRID and {\it Fermi}-LAT, the prompt {\it Swift}-XRT X-ray follow-up and simultaneous GASP-WEBT photometric and polarimetric optical data.
Second panel:
{\it Fermi}-LAT blue points from \cite{Ackermann2016}, red points from the public on line 
Fermi data analysis tool at SSDC.
In the last three panels we report a selection of the full dataset of
GASP-WEBT observations already presented in Figure \ref{GASP-WEBT}, zoomed 
around the $\gamma$-ray peak.}
\label{GRID-LAT-MW}
\end{figure*}
%
%
\section{RESULTS}

\subsection{Light curves}
\label{sect-lc}
In Figure \ref{GRID-GASP}, we present the simultaneous
(and as yet unpublished) AGILE $\gamma$-ray and GASP-WEBT optical
light curves during the 3C 279 flare in June 2015.
In order to produce the AGILE light curve, 
we divided the data  collected in the period from 11 to 18 June 2015
(MJD: 57184 -- 57191) in 24-hour and 12-hour timebins.
To derive the estimated flux of the source, we ran the
AGILE Multi-Source Maximum Likelihood
Analysis (ALIKE) task with an analysis radius of 10$^{\circ}$.
The ALIKE was carried out by fixing the position of the source to
its nominal radio position~\citep{1995AJ....110..880J},
(l, b) = (305.104, 57.062) (deg),
and using Galactic and isotropic diffuse emission parameters
(GAL-ISO) fixed at the values estimated during the two weeks
preceding the analyzed AGILE dataset.

The extended GASP-WEBT optical light-curve ($R$-band magnitude)
of 3C 279 since the end of 2014,
including the $\gamma$-ray flaring period
(MJD: 57010 -- 57220), is shown in Figure \ref{GASP-WEBT}.
It includes polarization percentage $P$ and electric vector polarization angle (EVPA) variations.
The total brightness variation in this period is $\sim 1.5$ magnitude,
from $R=16.07$ at MJD=57142.1 to $R=14.58$ at MJD=57189.6.

The multi-wavelength behavior of the source during the flare is then
summarized in Figure \ref{GRID-LAT-MW}, which includes
$\gamma$-ray light curves, as observed by AGILE-GRID and {\it Fermi}-LAT, the prompt {\it Swift}-XRT X-ray follow-up and simultaneous GASP-WEBT de-absorbed optical flux densities and polarimetric data.

A well-defined
maximum peaking around MJD=57189 is visible at $\gamma$-rays, in
agreement with the optical observations.
The degree of observed polarization $P$ remains always high, ranging
between about 9\% and 30\%.
The maximum observed value occurs
at MJD=57190.2, and the daily sampling allows to
identify a one-day delay of the $P$ maximum after the flux peak observed at optical and $\gamma$-ray frequencies.
The rise and the following decrease of $P$ and flux is
accompanied by a rotation of the electric vector polarization angle
of about 30\degr\ in 10 days.

As shown in Figure \ref{GRID-LAT-MW}, third panel, also the X-ray flux variability
appears correlated with the $\gamma$-ray and optical ones.
The peak X-ray flux value occurs at MJD=57189.14, and it is about a factor of
about 4 higher than the one observed one day later, see Table \ref{xspec}.

\subsection{Spectral Energy Distribution}
\label{sect-sed}
\begin{figure*}
 \centering
\vskip 0.2 true cm
\includegraphics[angle=-90,width=17cm]{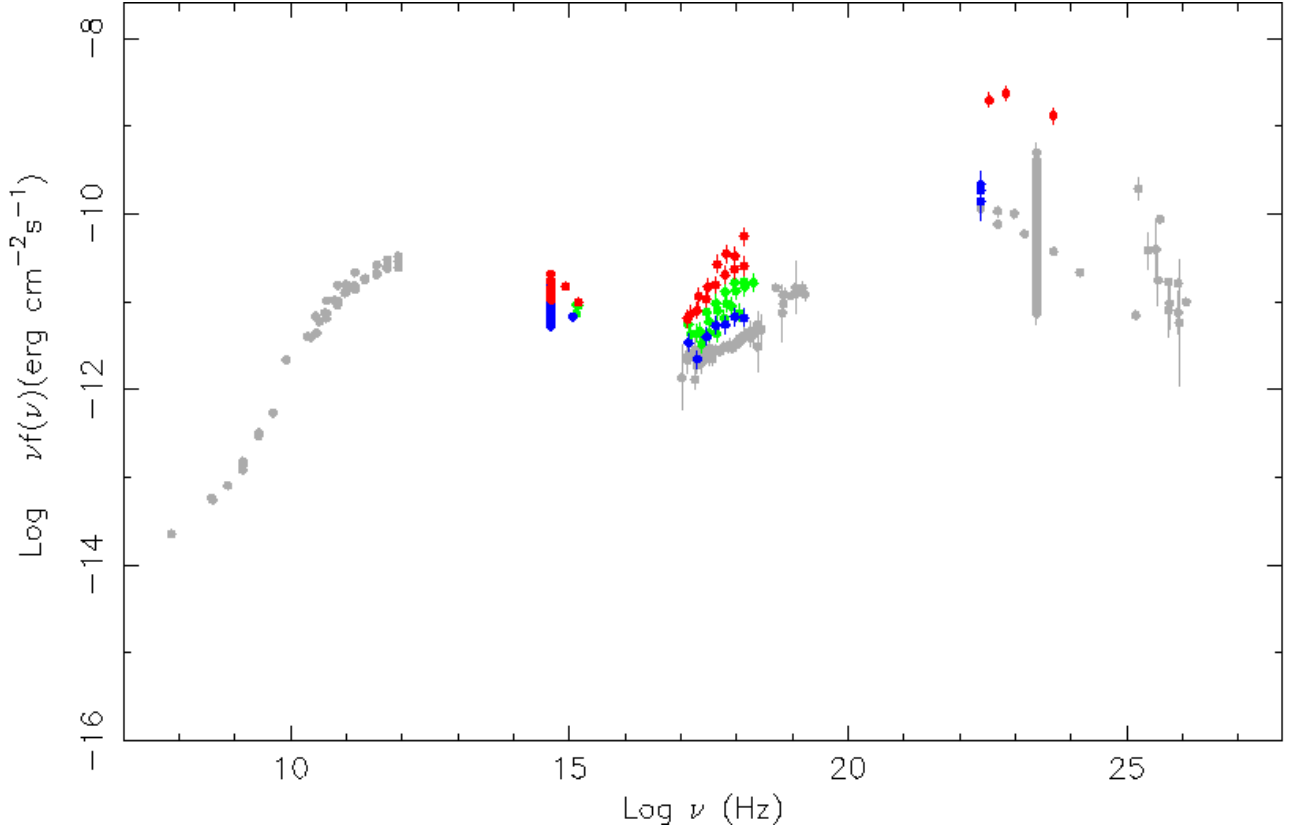}
\caption{The 3C 279 broad-band spectral energy distribution obtained with the help of the SSDC SED Builder tool (V3.2). {\bf Red points:} AGILE data during the June 2015 $\gamma$-ray flare (around MJD: 57187--57190), 
and simultaneous GASP-WEBT, {\it Swift}--UVOT and {\it Swift}--XRT ToO data. 
{\bf Green points:} {\it Swift}-UVOT and {\it Swift}-XRT follow-up data covering approximately 48 hours after the $\gamma$-ray peak emission (see Table \ref{swift-summary}). 
{\bf Blue points:} Post-flare 2015 data from GASP-WEBT (up to MJD=57220), {\it Swift}--UVOT, {\it Swift}--XRT (MJD=57191) and AGILE data (weekly averaged flux above 100 MeV from MJD=57197.5 to 57218.5).
{\bf Grey points:} public non-simultaneous archival data from SSDC (CRATES, DIXON, NVSS, PKSCAT90, PMN,
VLSS, AT20GCAT, PLANCK, WMAP5, {\it Swift}-BAT, IBIS/ISGRI, BeppoSAX, AGILE-GRID, {\it Fermi}-LAT, MAGIC).
}
\vskip 0.2 true cm
\label{SED}
\end{figure*}

Figure \ref{SED} shows the 3C 279 broad-band spectral energy distribution obtained with the help of the SSDC SED Builder tool\footnote{http://tools.asdc.asi.it/SED}.
Simultaneous AGILE, GASP-WEBT, {\it Swift}-XRT and {\it {\it Swift}}-UVOT data during the June 2015 flare are shown in red.
Average $\gamma$-ray flux excluding the flaring period, and other public non-simultaneous archival data in other wavelengths are shown in gray.

%
\begin{table*}[t!]
\centering
\begin{tabular}{ccccc}  
\hline
\hline
~~~ Label &  ~~~~~~~T$_{\mathrm{start}}$ & ~~~~~~~T$_{\mathrm{stop}}$  & F (E $\geq$ 100 MeV) & ~~~$\Gamma_{\gamma}$ \\  
 ~~~~~~~  &  ~~~~~~~MJD       & ~~~~~~~MJD       &  [$10^{-6}$ ph cm$^{-2}$ s$^{-1}$]   &              \\
\hline
Pre-outburst & ~~~~57184.0 & ~~~~~~ 57187.0  & ~~~ $(1.7 \pm 0.7)$  &  $ (2.0 \pm 0.4)$ \\
Flare        & ~~~~57187.0 & ~~~~~~ 57190.0  & ~~~~$(13.0 \pm 1.3)$    &  $ (2.1 \pm 0.1)$ \\  
Post-flare   & ~~~~57190.0 & ~~~~~~ 57193.0  & ~~~~$(1.0 \pm 0.5)$  &  -- \\
\hline
\hline
\end{tabular}
\caption{AGILE $\gamma$-ray fluxes and spectral indices.
Over the considered 3-day time periods, the source flux increases of factor of about 7,
then rapidly drops more than a factor of 10 in the post-flare, with insufficient statistics for spectral analysis.
}
\label{AGILE-fluxes}
\end{table*}

We have performed the AGILE spectral analysis of the peak $\gamma$-ray activity, corresponding to the period between 2015-06-14 (MJD=57187.0) and 2015-06-17 (MJD=57190.0) over three energy bins: 100--200, 200--400, and 400--1000 MeV.
A simple power-law spectral fitting gives a photon index of $ \Gamma_{\gamma}=(2.14 \pm 0.11)$,
consistent within the errors with the values reported by Fermi ~\citep{Ackermann2016,2015ApJ...803...15P}.
Moreover, we estimated the average $\gamma$-ray fluxes obtained by
integrating in the whole AGILE energy band (100 MeV -- 50 GeV)
during three time periods defined as {\it pre-outburst}
(MJD: 57184--57187), {\it flare} (MJD: 57187--57190), and {\it post-flare}
(MJD: 57190--57193). The corresponding AGILE integral $\gamma$-ray
fluxes and spectral indices are summarized in Table
\ref{AGILE-fluxes}. Historically this is the largest $\gamma$-ray
flare ($\geq 100$ MeV) of 3C 279 ever observed, including recent
activity reported in \cite{2017ATel10563....1B}.

The spectral energy distribution during the flare (red points in 
Figure \ref{SED}) indicates a very
high \lq \lq Compton dominance'': the ratio of the inverse Compton peak to the synchrotron one is of order $100$.
Specifically, the $\gamma$-ray spectrum integrated over 1-day timebins
rises by a factor of $\sim 3$ in a few hours (as shown in Figure \ref{GRID-LAT-MW}),
yielding a Compton dominance of about $100$, and attaining values up to $\sim 200$ when integrating on even shorter time-scales \citep{Ackermann2016}.

\section{Simple Flare Modeling and Discussion}
\label{sec:modeling}

In this section we estimate the parameters of a tentative 
simple modeling of the multi-wavelength 3C 279 
data acquired during the 2015 flare.
The model parameter values obtained here
can be used as reference input for more detailed 
further theoretical analysis.

In the framework of the one-zone leptonic model for FSRQs (see e.g.,
\citealt{Paggi2011}), the optical and UV data acquired during the
June 2015 flare, and presented here, would constrain the luminosity of
the accretion disk to $L_D\leq 10^{46}$~erg s$^{-1}$. We note that 
this value is larger by a factor of about 3 
than the disk luminosity previously inferred for
3C279 \citep{Raiteri2014}. 

Taking into account also the
simultaneous soft X-ray data and the observed variability, we can
determine empirical constraints on the model parameters: the
size $l$, the bulk boost factor $\Gamma$, the energetic content in
magnetic field $B$, and the electron energy distribution
$n_e(\gamma)$ of the emitting region. We assume that the
relativistic electrons have a double power law energy-density
distribution:
\begin{equation}
n_e(\gamma)=\frac{K\,\gamma_b^{-1}}{
(\gamma/\gamma_b)^{\zeta_{1}}+(\gamma/\gamma_b)^{\zeta_{2}}}\,[\rm
\,cm^{-3}], \label{eq-model}
\end{equation}
where $K$ is a normalization factor, $\gamma_b$ is the break Lorentz factor,
$\zeta_{1}$ and $\zeta_{2}$ are the double power-law spectral indices below and above the break, respectively.

\begin{figure*}
 \centering
\vskip 0.2 true cm
\includegraphics[angle=-90,width=14cm]{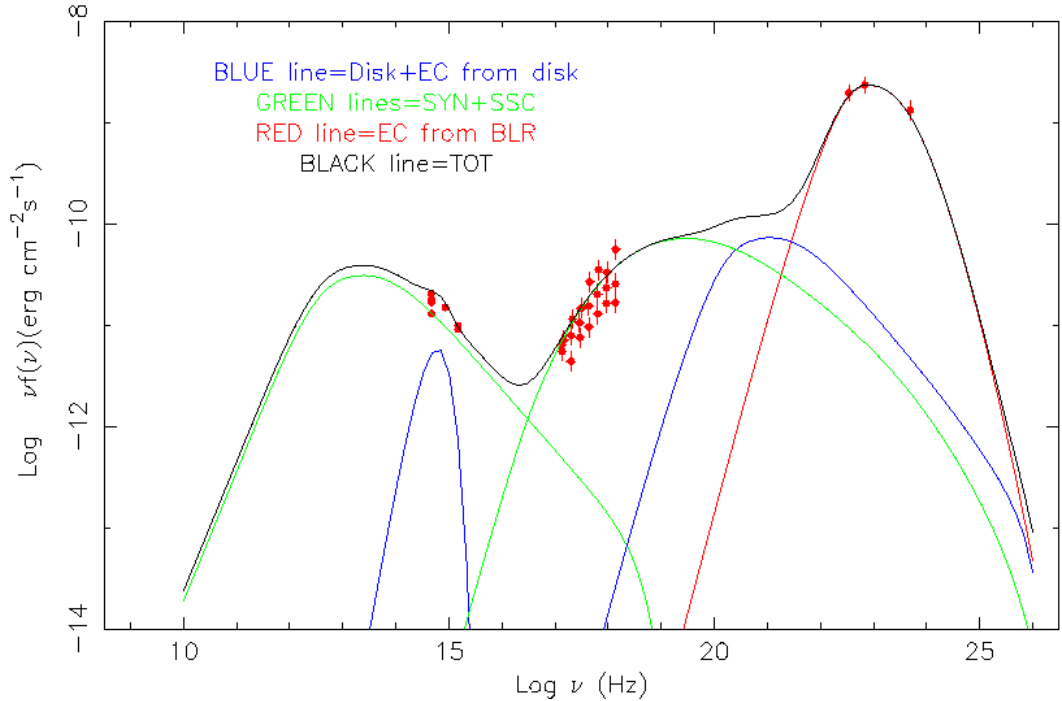}
\caption{3C 279: simple one-zone modeling of the June 2015 flare.
}
\label{SED_flare-model}
\end{figure*}

These electrons interact via the IC
process with the synchrotron photons internal to the same emitting region, and
with the external photons coming from the accretion disk and from the BLR.
The latter reflects
from distances $R_{\rm BLR}\simeq 0.1$ pc a fraction $\xi\simeq$ few $\%$ of
the disk radiation. 
In Figure \ref{SED_flare-model}
we show our one-zone SED model
of the June 2015 flare of 3C 279 for $\gamma$-ray fluxes averaged on 1-day timescales.
If we assume the emitting region located
at a distance $R<R_{\rm BLR}$ from the
central black hole, seed photons
coming from BLR are good candidates to be scattered into
$\gamma$-rays of observed energies
$\geq 100$ MeV, see red line in Figure \ref{SED_flare-model}.
As shown by blue lines in the same figure,
disk photons entering the emitting region from behind, are
scattered mainly in the hard X-ray observed band. Instead, the
internal scattering of the synchrotron photons are seen mainly in
the soft X-ray band, as shown by green lines.

In this model, we consider the emitting region placed at a
distance $R\,=\,6\, \times 10^{16}$ cm from the central black
hole, while the accretion disk radiates the power
$L_D\,=\,10^{46}$~erg s$^{-1}$: a fraction $\xi\,=\,2\%$ of this
is reflected back from the BLR placed at distance $R_{\rm
BLR}\,=\,0.15$ pc.
A summary of the best-fit flare model parameters are shown in
Table \ref{model-param}.

When the IC scattering occurs in the Thomson regime, the Compton dominance reads
$q\,=\,U'_{\rm ext}\,/\,U'_B$,
i.e. the ratio of the comoving energy density of BLR seed
photons $U'_{\rm ext}\simeq
(1+\beta_{\Gamma}^2)\,\Gamma^2\xi\,L_D\,/\,(4\pi\,c\,R_{\rm BLR}^{2})$
to the energy density of the magnetic field $U'_B=B'^2/8\pi$,  thus:
\begin{equation}\label{}
q\,\lesssim\,
0.2\,\Gamma^2\frac{(\xi/0.02)L_{D,46}}{(B'/{\rm G})^2(R_{\rm BLR}/0.1 \rm pc)^2}\,.
\end{equation}
For assumed disk luminosities $L_D\leq 10^{46}$~erg s$^{-1}$
this yields a value $q\,\leq\, 80$.
Moreover, the one-zone assumption has two other main consequences.
\begin{itemize}
\item First, a strict correlation of optical and $\gamma$-ray fluxes: their
variations must be of the same entity, so the Compton dominance should not vary.
\item
Second, to increase the upper limit for $q$
up to values above 100, as observed,
we should consider faint magnetic
fields values $B\,\lesssim\,0.1$ G, which would in turn imply modest electron accelerations \citep{Mignone2013}.
Alternatively, we could assume
bulk factors $\Gamma\,>\, 30$ \citep{Ackermann2016}, 
considerably exceeding the value $\Gamma\,\simeq\,20$ inferred from radio observations for this source \citep{Hovatta2009},
that would imply a conspicuous kinetic load in the jet.
\end{itemize}

Noticeably, the multi-wavelength light curves of the flare in Figure \ref{GRID-LAT-MW}
show instead that the Compton
dominance rises by a factor of three or more in a half day, attaining values up to
$q\,>\,200$ in few minutes when considering the very fast and
strong $\gamma$-ray variations reported in \citet{Ackermann2016}.
While the simple one-zone model presented here could account for the SED flaring
data integrated on 1-day timescales (provided you assume of a very bright 
underlying disk), it is anyway seriously challenged by the
observed strong and fast variation of the Compton dominance.

Furthermore, we notice that a single photon of energy $E = 52$ 
GeV was detected on MJD=57189.62 \citep{2015ApJ...803...15P} in
correspondence with the peak of optical emission and consistent
with the observed polarization fraction reaching its maximum.
Modelling of this specific episode of high-energy emission goes
beyond the scope of this paper, and provides an additional
argument for alternative modes of $\gamma$-ray emission. 
%

\vskip 0.5 cm
\begin{table}[t]
\centering
\begin{tabular}{|c|c|c|c|c|c|c|c|}
\hline
$\bf{l}$ & $\bf{K}$ & $\bf{\gamma_b}$ & $\bf{\gamma_{min}}$ & $\bf{\zeta_1}$ & $\bf{\zeta_2}$ & $\bf{\Gamma}$ & $\bf{B}$ \\
(cm) & (cm$^{-3}$) & & & & & & (G) \\
\hline
$10^{16}$ & 1100 & 700 & 180  & 2 & 4.2 & 20 & 1 \\
\hline
\end{tabular}
\caption{\noindent One-zone model parameters for the 2015 flare of 3C 279, as defined in Sect. \ref{sec:modeling}.}
\label{model-param}
\end{table}

\section{Conclusions}

In this paper we present multifrequency optical and X-ray
data simultaneous with the 2015 $\gamma$ flaring activity of 3C 279. 
We use AGILE-GRID and {\it
Fermi}-LAT~\citep{Ackermann2016,2015ApJ...803...15P} $\gamma$-ray
data together with
{\it Swift}-UVOT, {\it Swift}-XRT, and as yet unpublished optical
GASP-WEBT observations of 3C 279 in June 2015.
We find that from the multi wavelength light curve shown in
Figure \ref{GRID-LAT-MW}, 
the high-energy flare is partially correlated with the
behavior in other energy bands. Specifically, the $\gamma$-ray flux
rising by a factor $\simeq 4$ in half a day shows an optical
counterpart rising only by a factor 2 or less on similar time-scales.
The $\gamma$-ray flux during this flare exceeds the
largest 3C 279 flares previously detected, although
\cite{2015ApJ...807...79H}
reported an even more extreme multi-frequency 
behavior for this source in the past: e.g., in December 2013 the
$\gamma$-ray flux above $100$ MeV jumped by a factor $\simeq 5$ in a
few hours without optical or X-ray counterparts, and the Compton
dominance attained values of about 300. 
\cite{Ackermann2016} discuss variability of the 2015 $\gamma$-ray 
flare with minute timescales.

The observed spectral characteristics and the strong
and fast variations of the Compton dominance challenge  one-zone
models, unless we assume 
significant variations in the field of seed photons to be
IC scattered into $\gamma$-rays. We discuss in this paper a
one-zone model and provide the model parameters that can be used as
a theoretical model of reference. Models alternative to standard
SSC and EC might be considered 
(e.g. \citealt{Ackermann2016}).
In the moving mirror model  \citep{2015ApJ...814...51T,Vittorini2017}
localized enhancements of synchrotron photon density  
may explain the occurrence of gamma-ray flares with faint or no
counterpart in other bands. These localized enhancements would
persist only for short periods of time, and this would explain
the fact that the majority of FSRQ $\gamma$-ray flares are not
orphan in nature.

We noticed that, as shown in Figure \ref{GRID-LAT-MW}, the
degree of observed optical polarization $P$ appears to correlate with the optical flux 
$F$ during the flare, with $P$ peaking about
one day after $F$. Moreover, the polarization angle
rotates by at least 30\degr\ in the period encompassing the flare.
However, the behavior of the polarization degree of the jet may be very different
from the observed one, due to the big blue bump dilution effect.
When deriving the intrinsic jet polarization $P_{jet}$, the presence of a very luminous disc, as assumed by the one-zone model used to interpret the observed SEDs, would imply that the correction for the thermal emission contribution becomes noticeable as the flux approaches the observed minimum level. This would lead 
to much higher $P_{jet}$ values than the observed ones, and 
$P_{jet}$ would not maintain the general correlation with flux shown in
Figure \ref{GASP-WEBT}.


\vskip 0.5 cm

\section*{Acknowledgements}
Partly based on data taken and assembled by the WEBT collaboration
and stored in the WEBT archive at the 
Osservatorio Astrofisico di Torino - INAF\footnote{http://www.oato.inaf.it/blazars/webt/}. For questions about data availability contact the WEBT President, Massimo Villata (villata@oato.inaf.it).

We would like to acknowledge the financial support of ASI under contract to INAF: ASI 2014-049-R.0 dedicated to SSDC.
Part of this work is based on archival data, software or online services provided by the ASI Space Science Data Center (SSDC,
previously known as ASDC).
The research at Boston University was supported by National Science Foundation grant AST-1615796 and NASA Swift Guest Investigator
grant 80NSSC17K0309.
This research was partially supported by the Bulgarian National Science Fund of the Ministry of Education and Science under
grant DN 08-1/2016.
The Skinakas Observatory is a collaborative project of the University of Crete, the Foundation for Research and Technology -- Hellas,
and the Max-Planck-Institut f\"ur Extraterrestrische Physik.
St.Petersburg University team acknowledges support from Russian Science Foundation grant 17-12-01029.
This article is partly based on observations made with the telescope IAC80 operated by the Instituto de Astrofisica de Canarias
in the Spanish Observatorio del Teide on the island of Tenerife. The IAC team acknowledges the support from the group of support
astronomers and telescope operators of the Observatorio del Teide. Based (partly) on data obtained with the STELLA robotic
telescopes in Tenerife, an AIP facility jointly operated by AIP and IAC.
This work is partialy based upon observations carried out at the Observatorio Astron\'omico Nacional on the
Sierra San Pedro M\'artir (OAN-SPM), Baja California, Mexico.
C.P., V.V. and M.T. also thank Prof. A. Cavaliere for insightful discussion. 
{\bf 
\software{AGILE software Package (AGILE SW 5.0 SourceCode), XRTDAS (v.3.1.0), HEASoft package (v. 6.17), XSPEC}. 
}
\bibliographystyle{aasjournal}
\bibliography{AGILE_3C279_ApJ}

\label{lastpage}
\end{document}